\documentstyle{mn2e}
\input psfig

\def\solmas{{M$_\odot$}}
\def\simless{\mathbin{\lower 3pt\hbox
   {$\rlap{\raise 5pt\hbox{$\char'074$}}\mathchar"7218$}}}   
\def\simgreat{\mathbin{\lower 3pt\hbox
   {$\rlap{\raise 5pt\hbox{$\char'076$}}\mathchar"7218$}}}   
\def\etal{{\rm et al.}}

\def\solmas{{M$_\odot$}}
\def\solm{{M_\odot}}

\def\solr{{R_\odot}}

\def\tcoll {\tau_{\rm c}}
\def\Rcapt {R_{\rm capt}}
\def\Rcapt {R_{\rm t}}

  \makeatletter
  \ifx\CUP@mtlplain@loaded\undefined  
  \makeatother  
    
  \else
    
  \fi

\loadboldmathitalic 
\loadboldgreek      
  
\makeatletter
\ifx\CUP@mtlplain@loaded\undefined  
  \newfont\bit{cmbxti10 at 9pt}
\else
  \newfont\bit{mtbxti10 at 9pt}
\fi
\makeatother

\def\LaTeX{L\kern-.36em\raise.3ex\hbox{a}\kern-.15em
    T\kern-.1667em\lower.7ex\hbox{E}\kern-.125emX}

\newcommand{\gsim}{\mathrel{\hbox{\rlap{\lower.55ex \hbox {$\sim$}}
                   \kern-.3em \raise.4ex \hbox{$>$}}}}
\newcommand{\lsim}{\mathrel{\hbox{\rlap{\lower.55ex \hbox {$\sim$}}
                   \kern-.3em \raise.4ex \hbox{$<$}}}}

\title[Brown dwarfs in globular clusters] {Are there brown dwarfs in
globular clusters?}  \author[Bonnell
\etal]{I. A. Bonnell$^1$\thanks{E-mail: iab1@st-and.ac.uk},
C. J. Clarke$^2$, M. R. Bate$^3$, M. J. McCaughrean$^4$,
J. E. Pringle$^2$\\ \\ \LARGE and H. Zinnecker$^4$\\ 
$^1$ School of Physics and Astronomy, University of St Andrews, North Haugh, 
St Andrews, Fife, KY16 9SS. \\ 
$^2$ Institute of Astronomy, Madingley Road, Cambridge, CB3 0HA. \\ 
$^3$ School of Physics, University of Exeter, Stocker Road, Exeter EX4 4QL \\ 
$^4$ Astrophysikalisches Institut Potsdam, An der Sternwarte 16, 14482 
Potsdam, Germany }

\date{\today}
\begin{document}

\maketitle

\begin{abstract}


We present an analytical method for constraining the substellar
initial mass function in globular clusters, based on the observed
frequency of transit events. Globular clusters typically have very
high stellar densities where close encounters are relatively common,
and thus tidal capture can occur to form close binary
systems. Encounters between main sequence stars and lower-mass objects
can result in tidal capture if the mass ratio is $\simgreat
10^{-2}$. If brown dwarfs exist in significant numbers, they too will
be found in close binaries, and some fraction of their number should
be revealed as they transit their stellar companions. We calculate the
rate of tidal capture of brown dwarfs in both segregated and
unsegregated clusters, and find that the tidal capture is more likely
to occur over an initial relaxation time before equipartition
occurs. The lack of any such transits in recent HST monitoring of
47\,Tuc implies an upper limit on the frequency of brown dwarfs
($\simless 15$\% relative to stars) which is significantly below that
measured in the galactic field and young clusters.

\end{abstract}


\section{Introduction}
Brown dwarfs have recently been shown to exist in significant numbers
in the nearby galactic field (Nakajima \etal{} 1995; Rebolo,
Zapatero-Osorio, \& Martin 1995; Reid \etal{} 1999), young open
clusters such as the Pleiades, M\,35, and $\alpha$\,Per (Bouvier
\etal{} 1998; Barrado y Navascues \etal{} 2001, 2002) and star-forming
regions including the Orion Trapezium Cluster, Taurus-Auriga,
$\rho$\,Ophiuchi (McCaughrean \etal{} 1995; Hillenbrand \& Carpenter
2000; Luhman \etal{} 2000; Muench \etal{} 2002; Brice\~no \etal{}
2002; Luhman \& Rieke 1999; Muench \etal{} 2003; Preibisch, Stanke, \&
Zinnecker 2003). Analyses show that brown dwarfs may well be as
frequent as stars (Chabrier 2002), and thus although they are
insignificant in terms of their mass contribution, their existence and
numbers pose significant constraints on theories for the origin of the
Initial Mass Function (IMF).

One standard theory for the origin of the IMF is that the masses are
determined by the local thermal Jeans mass, such that brown dwarfs
form from very dense and cool regions of giant molecular clouds (see,
e.g., Padoan \etal{} 2001). Another possibility is that stellar masses
are determined through a combination of the local thermal Jeans mass
coupled with dynamical processes during the star formation process
(e.g.\ Bonnell \etal{} 1997; Klessen, Burkert \& Bate 1998; Bonnell
\etal{} 2001; Larson 2002; Bate, Bonnell \& Bromm 2003). In this
scenario, the low-mass end of the IMF arises due to dynamical
interactions that eject forming stars from their natal environments,
thus truncating their accretion and limiting their mass (Reipurth \&
Clarke~2001). A recent numerical simulation has shown that this
process can occur in dense stellar environments, such that objects
that would otherwise accrete sufficiently to become stars are reduced
to being brown dwarfs (Bate, Bonnell, \& Bromm 2002). Establishing the
low-mass end of the IMF in differing environments is therefore
important, in order to establish whether or not the IMF is universal,
and if not, which physical parameters play the dominant role in
defining its form. In particular, globular clusters may yield
important clues, as their low-metallicity environment should result in
lower cooling rates and thus larger Jeans masses (e.g. Larson~1998). In
fact, observations of the low-mass stellar component in globular
clusters yields IMFs similar to young stellar clusters and the
Galactic field (Paresce \& De Marchi~2000). This would imply that
globulars should have similar numbers of brown dwarfs as have been
found in nearby regions.

In this letter, we discuss an analytical method for constraining the
brown dwarf population of a globular cluster, based on observations of
transits.  In Section~2, we show that if significant numbers of brown
dwarfs are present in a globular cluster, tidal capture would lead to
a population of close binary systems comprising a normal main sequence
star and a brown dwarf, and that some fraction of these systems should
then exhibit transits. In Section~3, we examine how the frequency of
transit events in a given cluster can be used to quantify the imposed
constraint on the substellar mass function in the cluster. In
Section~4, we look at the results of a recent HST monitoring study of
47\,Tuc searching for planetary transits (Gilliland \etal{} 2000), and
argue that the complete lack of transits in their data implies a
dearth of brown dwarfs as close companions.

\section{Tidal capture}

Tidal capture occurs when two objects pass close enough together to
raise tides with energy equivalent to the excess kinetic energy of
their mutual orbits. In order for the capture mechanism to operate,
the gravitational perturbation from this passage should be large,
implying in turn that the mass ratio of the two stars should not be
too small: very low-mass perturbers do not raise large tides. An
additional requirement is that both objects should be single, as hard
binaries would repel any perturbers. Finally, to be efficient in a
statistical sense, the frequency of close interactions should be
significant, as is easily satisfied in a globular cluster.

Tidal capture was initially conjectured as the source of low-mass
X-ray binary systems found in globular clusters (Fabian, Pringle \&
Rees 1975).  The energy raised in a tide (Press \& Teukolsky 1977; Lee
\& Ostriker 1986; Fabian \etal{} 1975) by an object of mass $M_p$
passing within a distance $R$ of a star of mass $M_*$ and radius $R_*$
is
\begin{equation}
E_{ \rm tide} \approx {G M_*^2 \over R_*} \left({M_p \over M_*}\right)^2 
                      \left({R_*\over R}\right)^{6},
\end{equation} 
where $G$ is the gravitational constant. The detailed mode analysis of
Lee \& Ostriker (1986) showed that, for $n=3/2$ polytropes appropriate
for low-mass main-sequence stars, including the roughly equal
contributions from the quadrupole and octopole terms results in a
similar expression for the tidal energy. Here we consider only the
tides raised on the primary by a low-mass perturber, as the smaller
radius of the secondary would necessitate a direct collision before
sufficient energy was raised in its tides.

\begin{figure}
\centerline{\psfig{figure=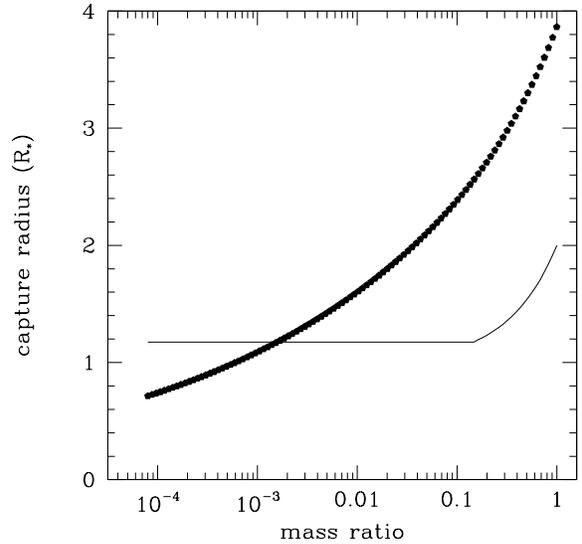,width=3.5truein,height=3.50truein}}
\caption{\label{rcapt} The capture distance $a_{\rm c}$ is plotted
(pentagons) in units of the stellar radius, as a function of the mass
ratio of the encounter. We have taken the mass and radius of the star
to be $0.5 \solm$ and $0.57\solr$, respectively. The collisional
radius is shown as a continuous line. Mergers (as opposed to capture)
will result for encounters where $a_{\rm c}< R_{\rm c}$. The rate of
tidal capture is proportional to the difference between the two lines
(see text). Interactions between stellar and substellar objects with
mass ratios greater than $\approx 10^{-2}$ have a significant chance
of resulting in a tidal capture.}
\end{figure}

Comparing this tidal energy to the excess kinetic energy over that of
a bound orbit at a separation of $R$, Fabian \etal{} (1975) showed
that in order for the star to capture the perturber, it must pass
within a periastron distance $R=a_{\rm c}$ of the star,
\begin{equation}
{a_{\rm c} \over R_*} = 
     \left({G M_* \over R_* v_{\rm d}^2}q (1+q)\right)^{1\over 6},
\end{equation}
where $v_{\rm d}$ is the velocity dispersion of the system and
$q=M_p/M_*$ is the mass ratio (see also Lee \& Ostriker 1986). It is
evident from these equations that if the mass ratio of the encounter
is too extreme, the tidal energy is very low and the capture radius
$a_{\rm c}$ becomes comparable to or smaller than the stellar radius
$R_*$. This effect is illustrated in Figure~\ref{rcapt}, which plots
the capture radius as a function of the mass ratio, assuming a fixed
primary star with mass $M_* = 0.5\solm$ and radius $R_* = 0.57 \solr$
(i.e.\ assuming $R \propto M^{0.8}$ for main sequence stars), and a
cluster velocity dispersion of $v_{\rm d}$=10\,km\,s$^{-1}$.  We see
that as the mass ratio of the encounter decreases, the capture radius
decreases dramatically until, for $q \le 10^{-3}$, the capture radius
becomes smaller than the stellar radius. For example, the capture
radius of a mass ratio of $q=0.1$ is $a_c \approx 2.4 R_*$ while for a
mass ratio of $q=0.02$, the capture radius decreases to $a_c \approx
1.8 R_*$.

In more detail, the capture radius in fact needs to exceed the sum of
the radii of the two interacting objects, i.e.\ the distance at which
the two sources would collide. For our fiducial $0.5\solm$ primary,
perturbers with $q \ge 0.16$ will be stellar (i.e.\ $\ge 0.08\solm$),
and we use
\begin{equation}
R_p \approx q^{0.9}R_*~,
\end{equation}
while for lower-mass, brown dwarf perturbers, we use
\begin{equation}
R_p \approx 0.1 \solr,
\end{equation}
as the radius is roughly independent of mass for these low-mass objects 
supported by electron degeneracy pressure. Figure~\ref{rcapt} then also 
plots the collisional radius 
\begin{equation}
R_{\rm c} = R_* + R_p.
\end{equation}

The cross section for tidal capture (excluding collisions) is, in the
gravitationally focused regime, simply proportional to the difference
between $a_{\rm c}$ and $R_{\rm c}$ (see Section~3). The essential
result from Figure~\ref{rcapt} is that perturbers with mass ratios of
$q\simgreat 10^{-2}$ have a non-negligible chance of being tidally
captured if they undergo a close encounter with a low-mass main
sequence star in a globular cluster environment.

As a check, we can compare this capture radius to that found in
numerical simulations of encounters between objects with unequal
masses. Encounters in a globular cluster with $v_{\rm d} \approx
10$\,km\,s$^{-1}$ are well approximated as having a low relative
velocity at infinity. In this situation, Benz \& Hills (1992) (for
$q=0.2$) and Lai, Rasio, \& Shapiro~(1993) ($q=0.1$) found that
capture occurred when the two stars pass within twice the sum of their
radii. This is in good agreement with Figure~\ref{rcapt} where, for
mass ratios of $q\approx 0.1$, $a_{\rm c} \approx 2 R_*$.

In the following analysis, we assume that the typical star in a
globular cluster has a mass and radius of $0.5\solm$ and $0.57\solr$,
respectively, while any brown dwarfs present have fiducial masses and
radii of $0.05\solm$ and $0.1\solr$, respectively. 
Once a brown dwarf
is captured to form a binary, the system will circularise on a
relatively short timescale (Mardling~1996).  In the case of a high
eccentricity binary circularising at constant angular momentum, the
resultant binary separation is then $\approx 2 a_{\rm c}$
(Mardling~1996). Thus we take the mean binary parameters to be a
separation of $2 a_{\rm c} \approx 4 R_* \approx 2 \solr$, and an
orbital period of $\approx 14$ hours.

In our approach, we are implicitly assuming that the deposition of the
tidal energy has little effect on the primary's structure. This is
justified in that the total energy dumped into tides is but a small
(2.5 \%) fraction of the binding energy of the star and that most of
this is in the form of bulk motion of the star's spin. The deposition
of this energy is sufficiently deep in a $n=3/2$ polytope and the
damping sufficiently long such as to have minimal effect on our result
(Ray, Kembhavi \& Antia~1987; Kochanek~1992).

\section{Close binaries with brown dwarf secondaries}
Next, we need to calculate the expected number of tidally-captured
close binary systems containing brown dwarfs from the total population
of brown dwarfs present in the globular cluster. Including
gravitational focussing the timescale, $\tcoll$, for a star of mass
$M_*$ to pass within a distance $R$ of another star is found from
(Binney \& Tremaine~1987)
\begin{equation}
{1 \over \tau} = 16 \sqrt{\pi} n(r) v_{\rm disp} R^2 
                 \left(1+ {GM_*\over 2 v_{\rm d}^2 R}\right)~,
                 \label{eqn:tcoll}
\end{equation}
where $n(r)$ is the density of stars in the cluster as a function of
the distance $r$ from the cluster centre. For a globular cluster with
a velocity dispersion of $v_{\rm d} \approx 10$ km s$^{-1}$ (e.g.\ 47
Tuc; Gebhardt \etal{} 1995), the second (gravitationally focussed)
term of the parentheses is much larger than unity and thus dominates,
and the rate is then proportional to $R$ rather than $R^2$. Thus, the
rate of encounters that result in capture (but {\it not} in
collisions) is proportional to $a_{\rm c} - R_{\rm c}$, which we
henceforth define as $\Rcapt$. We can then write
Equation~(\ref{eqn:tcoll}) as
\begin{equation}
\tcoll \approx 7 \times 10^{10}~{10^5 {\rm pc}^{-3} \over n(r)}~
               {v_{\rm d} \over 10 {\rm km\ s}^{-1}}~{\solr \over \Rcapt}~
               {\solm \over M_*}~{\rm years}.
\end{equation}
Finally, if we assume the age of the cluster, ${t_{\rm age}}$, to be 
$\approx 10^{10}$ years, then $\tcoll \gg t_{\rm age}$, and we can then 
estimate the probability of capture as
\begin{equation}
P_{\rm capt} = {t_{\rm age} \over \tcoll}~.
\label{eqn:pcapt}
\end{equation}

A difficulty arises in that, in addition to the actual numbers of
brown dwarfs, we do not know their spatial distribution $n(r)$. At one
extreme, there is the possibility that the brown dwarfs simply follow
the stellar distribution. However, it is much more likely that the
cluster is relaxed and that there is mass segregation, with the other
extreme scenario being that the brown dwarfs are in a state of energy
equipartition with the stars.

\begin{figure}
%
\centerline{\psfig{figure=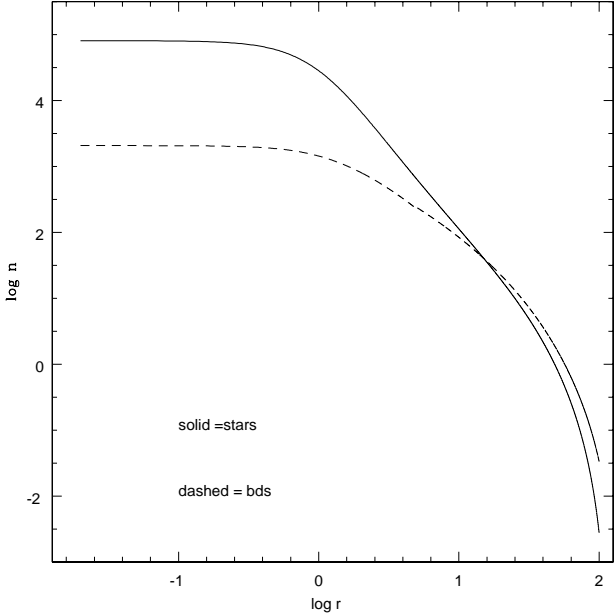,width=3.5truein,height=3.50truein,%
rwidth=3.5truein,rheight=3.5truein}}
\caption{\label{kingw9} The radial density distribution of stars (in
stars pc$^{-3}$) is plotted using a King model with $W=9$ (solid
line).  The density is normalised to the core density of 47 Tuc
(Djorgovski \& Meylan 1993). The distribution of brown dwarfs
following energy equipartition is also plotted (dashed line), assuming
equal numbers of brown dwarfs and stars in the cluster.}
\end{figure}

These two extremes are illustrated in Figure~\ref{kingw9}. The solid
line represents the W=9 King profile (or concentration level of
$c=log(R_{\rm tide}/ R_{\rm core})\approx 2$; Binney \& Tremaine
1987), appropriate for 47 Tuc (Trager, King, \& Djorgovski 1995),
assuming an equal number of brown dwarfs and stars in the cluster, and
normalising to a central stellar number density of $10^5$ pc$^{-3}$
(Webbink 1985). For the first of our extremes, the brown dwarf
distribution would follow this curve too.

The dashed line is then the brown dwarf distribution under the assumption
of equipartition of energy, computed from a multi-mass King model (da Costa 
\& Freeman 1976). We have simplified the construction of this model by 
considering just two mass bins (`stars' and `brown dwarfs'), and by neglecting 
the contribution of the latter population to the cluster potential. Since the 
brown dwarfs are on average about an order of magnitude less massive than the 
stars, we assign a $\sigma$ (the 1D velocity dispersion) in the King 
distribution function for the brown dwarfs that is three times the 
corresponding $\sigma$ for the stars. Again, we normalise the resulting
brown dwarf profile in Figure~\ref{kingw9} such that the total number of 
brown dwarfs is equal to the total number of stars. As expected, the 
equipartition model is significantly depleted in brown dwarfs in its inner 
regions.

Pending N-body models of globular clusters that include a realistic
number of stars and binaries, it is unclear which of the profiles
shown in Figure~\ref{kingw9} is more appropriate over the balance of
the cluster lifetime. To date, Monte Carlo simulations (which omit
binaries; Fregeau \etal{} 2002) and N-body simulations of less
populous clusters (Hurley \& Shara 2002) show a marked decrease in the
number of low-mass objects in the cores of clusters. The segregation
occurs over a number of relaxation times, but does not appear to reach
the level of equipartition. Our two extreme scenarios are therefore
likely to bracket the actual brown dwarf density profile. What remains
undetermined is the total number of brown dwarfs in globular clusters,
i.e., the normalisation of the distribution.

Proceeding, we next estimate the capture rate using the stellar distribution
plotted in Figure~\ref{kingw9}, sampling stars that appear (in projection) 
to be located between the centre of the cluster and the half-mass radius, 
as was the case for the HST observations of 47 Tuc (Gilliland \etal{} 2000). 

For each of the stars in the sample, we select orbital parameters from
the King distribution function and follow their orbits in the fixed King 
potential. At each instantaneous location of each sampled star, we compute 
a capture rate based on Equation~(\ref{eqn:pcapt}) and an assumed brown dwarf 
profile, whose shape is given by either of the extremes shown in 
Figure~\ref{kingw9}, and whose normalisation is adjustable. In this way, we 
accumulate a probability that the star has captured a brown dwarf, for both
assumptions about the (time invariant) brown dwarf distribution.

\begin{figure}
%
\centerline{\psfig{figure=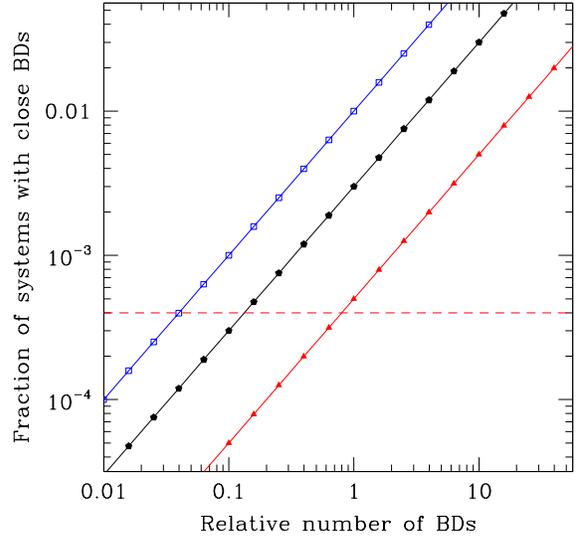,width=3.5truein,height=3.50truein}}
%
\caption{\label{fracBDs} The fraction of systems with brown dwarf
companions as a result of tidal capture plotted against the relative
number of brown dwarfs to stars in the cluster. The upper curve (open
squares) assumes that the brown dwarfs follow the same density
distribution as the stars over the full lifetime of the cluster. The
middle curve (pentagons) shows the fraction of systems captured within
an initial cluster relaxation time ($t_{\rm rh} \approx 3 \times
10^9$\,yrs; Harris 1996), before mass segregation becomes
important. The lower curve (filled triangles) shows the capture
probability assuming that the brown dwarfs are always in equipartition
with the stars over the cluster lifetime. The dashed line shows the
upper limit of $4 \times 10^{-4}$ stars with close brown dwarf
companions implied by the lack of transits seen in the 47 Tuc data of
Gilliland \etal{} (2000) (see text).}
\end{figure}

Figure~\ref{fracBDs} then shows the resulting fraction of stars with a 
tidally-captured brown dwarf companion, as a function of the ratio of the 
total number of brown dwarfs to stars in the cluster. The open squares 
correspond to the assumption that the brown dwarf density distribution follows 
that of the stars over the entire lifetime of the cluster, while the filled 
triangles illustrate the case where the brown dwarfs follow an equipartition 
distribution over the entire lifetime. Predictably, the capture rates are much 
lower in the latter case. Lastly, the filled pentagons correspond to the case 
where the brown dwarf distribution follows the stars for the first relaxation 
time of the cluster ($\approx 3 \times 10^9$ years; Harris 1996), and that 
the capture rate is negligible thereafter, as justified by the lower 
capture probabilities seen once equipartition is achieved.

For example, assuming that the stars and brown dwarfs follow the same
distribution and are equal in number, then 1\% of the stars should
have brown dwarf companions after $10^{10}$\,yrs. Conversely, for
equal numbers of brown dwarfs and stars in a fully segregated
distribution, roughly only 0.05\% of the stars should have brown dwarf
companions. Finally, for the scenario with a cluster equal in number
of stars and brown dwarfs, unsegregated for an initial relaxation time
of $t_{\rm rh} \approx 3 \times 10^9$\ yrs, and segregated thereafter,
$\approx 0.3$\% the stars end up with brown dwarf companions.

The above estimates neglect the possibility that the cluster is
undergoing core collapse. In such a scenario, the core density 
would have increased in the recent past and thus the core number densities
throughout much of the lifetime of the cluster would have been smaller.
This scenario appears unlikely considering the relatively low concentration
level of the density profile and the size of the core (Howell, Guhathakurta \& Gilliland~2000). We can make a crude estimate of such a scenario
by taking  the segregated distribution of the brown dwarfs as being the initial
distribution for both stars and brown dwarfs. This results, as above, 
in only 0.05\% of the stars having brown dwarf
companions.

\section{The case of 47 Tuc}
Gilliland \etal{} (2000) used the Hubble Space Telescope to monitor the 
globular cluster 47~Tuc continuously over 8.3 days, obtaining imaging 
photometry every few minutes, searching for transits. The HST WFPC2 
field-of-view spanned a region from the core out to the half-mass radius of 
the cluster and contained $\approx 34000$ stars. No transit events were 
detected during the monitoring campaign. 

The goal of the campaign was to detect hot Jupiters (i.e.\ in close orbits, 
cf.\ 51~Pegasi \etal) in orbits of 2--4 days around main sequence stars in 
the cluster. However, if present as close companions, brown dwarfs would
also have been readily detectable via transits. The radius of a brown dwarf
is essentially the same as that of Jupiter, so the fraction of the primary 
stellar surface covered would be similar. Also, at ages of $10^{10}$ years, 
even the most massive brown dwarfs should have cooled to temperatures of less 
than 1000\,K (Burrows \etal{} 1997), rendering them effectively black when 
viewed against a stellar disk.

The probability of a transit in a given system depends on $\sin(i)$, the 
orbital inclination with respect to the line-of-sight, and the orbital 
separation, with a transit requiring $\sin(i) < (R_* + R_{\rm BD}) / a$. 
For the hot Jupiters hoped for by Gilliland
\etal{} (2000), typical orbital parameters would have predicted a 10\%
probability that any given system would be inclined such that the planet 
would transit the stellar disk, leading Gilliland \etal{} (2000) to 
conclude that $\simless 10^{-3}$ of the stars in 47 Tuc have hot Jupiters. 
This fraction is significantly lower than seen for nearby galactic field 
stars, and one plausible explanation is that the low metallicity in globular 
clusters makes planet formation there a very inefficient process.

For tidally-captured brown dwarfs, however, the orbital separation
would be just $\approx 4 R_*$ orbit, and thus the chance of a transit
in a given system rises to 25\%. Thus, the same null result can be
used to infer that $\simless 4 \times 10^{-4}$ of the stars in 47 Tuc
have close brown dwarf companions today.
Therefore, in the unlikely case where the stars and brown dwarfs are in
equipartition for the full lifetime of the cluster, Figure~\ref{fracBDs} shows that
the total number of brown dwarfs in 47 Tuc must be less than the number
of stars. For the more probable scenario, where the cluster is
initially unsegregated for the first relaxation time, Figure~\ref{fracBDs} implies
that the number of brown dwarfs is $\simless 15$\% of the number of
stars.

Our fiducial calculations presented above assumed $0.5\solm$ stars
with $0.57\solr$ radii, while in their analysis of 47 Tuc, Gilliland
\etal{} (2000) assumed typical parameters of $0.81\solm$ and\
$0.92\solr$ for the stars in their 47 Tuc sample. For the sake of
consistency, it is thus worth estimating the difference their
assumptions would have on our calculations. The higher-mass and
larger-radius stars have higher rates of tidal capture, and working
through our analysis again, we find the expected fraction of stars
with tidally-captured brown dwarf secondaries to be a factor 2 higher
than estimated for our fiducial case. This translates into an increase
by a factor 2 for the predicted fraction of systems with close brown
dwarf companions: in the scenario where the cluster contains equal
numbers of brown dwarfs and stars, and where captures occur only
during an initial relaxation time when the brown dwarfs follow the
stellar distribution, the expected fraction of stars with brown dwarf
companions would increase from $0.3$\% to $0.6$\%, and the expected
number of transits would double. Thus, the absence of transits in 47
Tuc would imply a factor of 2 lower brown dwarf frequency, i.e.\
$\simless$7.5\% of the stellar frequency.

It is also worth considering what would happen if the typical brown
dwarf mass was decreased from $0.05$\solmas{} to 0.01\solmas, with a
constant radius of $\sim0.1\solr$. The cross section for capture
reduces to $\approx 60$\% of
its former value and thus the fraction of stars with brown dwarf
companions
then decreases from $0.3$\% to $0.18$\%. The number of expected
transits would decrease by 60\%, and thus the lack of transits would
imply a 60\% higher upper limit on the number of brown dwarfs than in
our fiducial case, i.e.\ the total brown dwarf population is
constrained to be $\simless$25\% of the stellar population. However,
even though the total number of brown dwarfs has gone up, their
contribution to the total cluster mass remains negligible.

In conclusion, taking the most plausible model for the dynamical
evolution of 47 Tuc and thus of the brown dwarf capture rates, the lack
of any detected transits in the cluster implies a significantly lower
ratio of brown dwarfs to stars ($\simless 15$\%) than has been found in
nearby clusters and the solar neighbourhood ($\sim$\,unity). This in
turn implies that the initial mass function is {\em not\/} in fact
universally constant below the hydrogen burning limit.

\section*{Acknowledgments}
We thank the Institute of Astronomy for being our host, and Steinn Sigurdsson 
for useful discussions. This work was partly funded by the European Commission Research
Training Network ``The Formation and Evolution of Young Stellar
Clusters'' (HPRN-CT-2000-00155).

\end{document}